\documentclass[showpacs,twocolumn,prb,aps,floatfix]{revtex4}
\usepackage{amsmath,amssymb}
\usepackage{graphicx,epsfig, color}
\newlength{\upit}\upit=0.1truein

\newcommand{\ltappr}{{{\lower4pt\hbox{$<$} } \atop \widetilde{ \ \ \ }}}
\newlength{\bxwidth}\bxwidth=1.5 truein

\newlength{\figwidth}
\newlength{\shift}
\newcommand{\fg}[3]
{
\begin{figure}[ht]

\vspace*{-0cm}
\[
\includegraphics[width=\figwidth]{#1}
\]
\vskip -0.2cm
\caption{\label{#2}
\small#3
}
\end{figure}}

\newcommand \bea {\begin{eqnarray} }
\newcommand \eea {\end{eqnarray}}

\newcommand{\br}{{\bf{r}}}

\begin{document}

\pacs{67.85.-d, 67.85.Fg, 71.10.Fd}
\title{Itinerant Ferromagnetism in an Atom Trap.}

\author{Ilya Berdnikov$^{1,2}$, P. Coleman$^1$ and Steven H. Simon$^2$}
\affiliation{$^{1}$ Center for Materials Theory, Rutgers University, Piscataway, New Jersey 08854, USA}
\affiliation{$^{2}$ Alcatel-Lucent, Bell Labs, Murray Hill, New Jersey 07974, USA}
\date{\today}

\begin{abstract}
  We propose an experiment to explore the magnetic phase transitions
  in interacting fermionic Hubbard systems, and describe how to obtain
  the ferromagnetic phase diagram of itinerant electron systems from
  these observations. In addition signatures of ferromagnetic
  correlations in the observed ground states are found: for large trap
  radii (trap radius $R_T > 4$, in units of coherence length $\xi$),
  ground states are topological in nature --- a ``skyrmion'' in 2D,
  and a ``hedgehog'' in 3D.
\end{abstract}

\maketitle
\section{Introduction}\label{}\label{}

The simplest, and best studied model of itinerant ferromagnetism (FM)
is the Hubbard model \cite{hubbard1963}. Shortly after its
introduction, Nagaoka and Thouless proved \cite{nagaoka1966,
  thouless1965} that for an interaction $U$ of infinite strength,
doping one hole into a background of spins leads to FM. Yet for more
than forty years, the fate of the Thouless-Nagaoka phase in the
Hubbard model at finite doping and finite interaction has not been
fully resolved.  Existing studies include various perturbative
\cite{oles1982, ioffe1988, putikka1992}, variational
\cite{shastry1990, linden1991, hanisch1993, wurth1996}, slave boson
\cite{moeller1993, boies1995}, Quantum Monte-Carlo \cite{zhang1991,
  becca2001} and DMFT \cite{park2008} calculations. For example, there
is no consensus as to the values of the critical doping $\delta_{cr}$
($\approx 0.19 - 0.49$) below which FM occurs and critical interaction
$U_{cr}$ ($\approx 63-77.7$ in units of the hopping amplitude $t$).
Moreover, the various approaches do not agree on the nature of the
transition, first order claimed by some\cite{oles1982,
  ioffe1988,moeller1993} second order by others\cite{zhang1991,
  becca2001, park2008}.

Experiments on optical lattices, which allow a tunable control of
model parameters, offer an interesting opportunity to address these
long-standing open questions. This setting was used to study
correlations in cold bosonic systems experimentally,
\cite{greiner2002,diener2007}, and could also be applied to
superconductivity of fermions. The superconducting transition,
however, requires a low entropy state cooled to temperatures far below
degeneracy, which still poses a significant challenge. Ferromagnetism
offers a particular advantage in this respect, since, at least for the
Nagaoka phase at $U=\infty$, the absence of any other scales sets the
transition temperature to be a finite fraction of the bandwidth, and
entropy at the transition is not a small parameter (of order unity per
carrier). A recent experiment on two-component fermionic ${}^{40}$K in
optical lattices has reported a Mott insulator with a maximum achieved
$U/t \sim 40$, $T/T_F=0.28$ and initial densities of more than one
electron per site\cite{joerdens2008}. A FM phase presumably resides
near the observed Mott phase, and conservative estimates place it at
slightly less than one particle per site, $U/t \sim 100$ and $T/T_F
\sim 0.1$\cite{hanisch1993,wurth1996,park2008}. The experimental
parameters required in a setup such as \cite{joerdens2008} are
estimated: lattice depth $V_0/E_r \sim 13-19$, depending on the
scattering length $a_s$, the coherence length $\xi
=\frac{\hbar}{\sqrt{2 m\mu}} \sim 0.2 \mu{}m$, and the trap radius
$R_T \sim 1.6 \mu{}m$, obtained from the laser waist radius of
$160\mu{}m$.

In this paper we explore the possibility of using cold fermion gases
to study FM in both 2D and 3D systems. For large trap sizes (radius
$R_T > 4$ coherence lengths $\xi$) with the constraint of zero total
magnetization and filling factor less than unity everywhere in the
trap, we find that the ``skyrmion'' configuration (Fig.
\ref{hedgeskyrm} (a)) has the lowest energy in 2D, while in 3D, the
lowest energy configuration is the ``hedgehog'' (Fig.
\ref{hedgeskyrm} (b)). Observing these ground states provides a simple
way to map out the FM phase diagram, which would be the
first result of its kind.

\figwidth=7cm
\fg{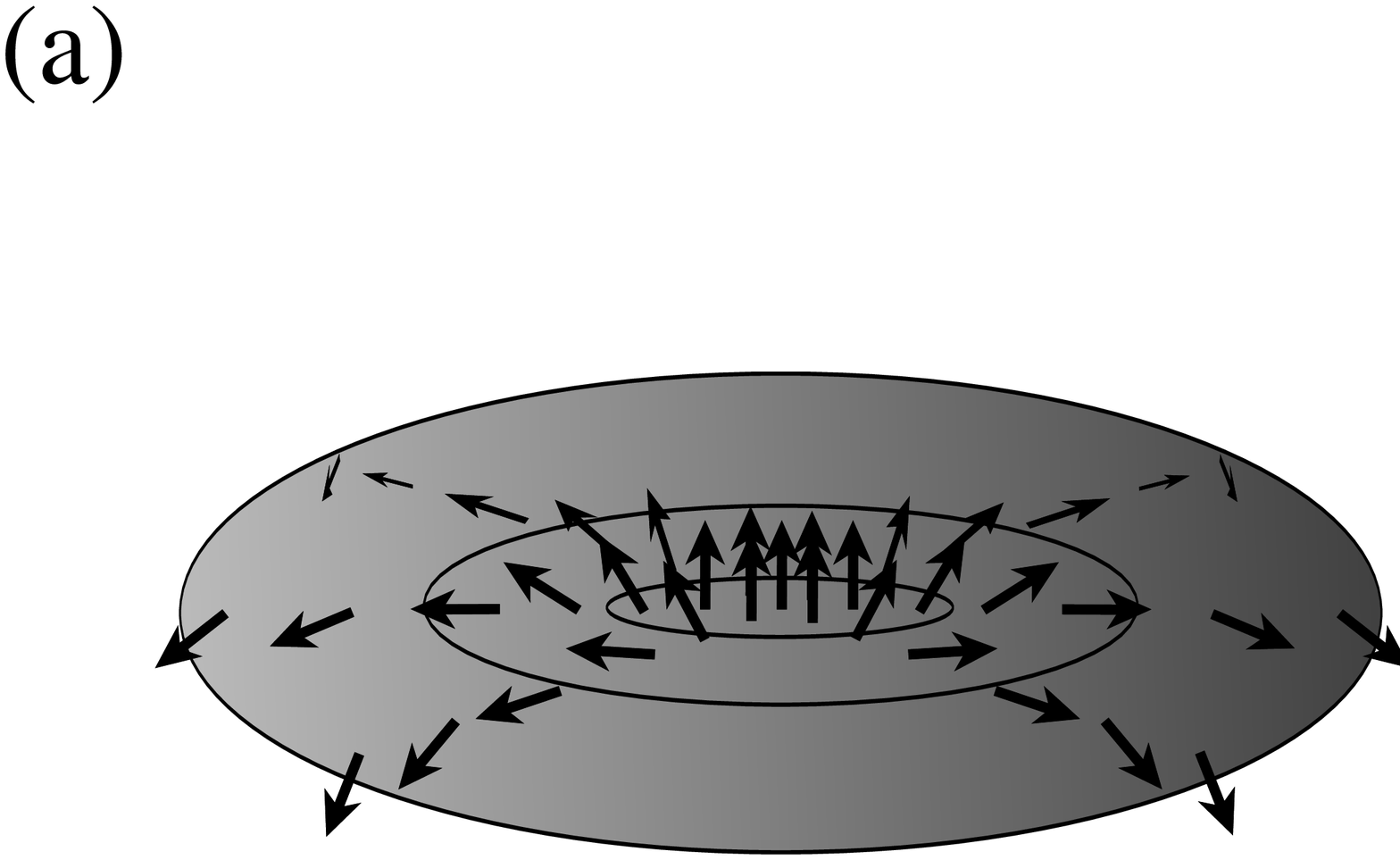}{hedgeskyrm}{(Color online) (a)Skyrmion and 
 (b) Hedgehog configuration of spins.}
%

\section{Approach}\label{}

Our approach to the problem is in marked contrast with previous
studies \cite{salasnich2000, sogo2002, duine2005,wu2009}, some of which
address only the 2D case: rather than using cold atoms as a means to
verify microscopic calculations, we propose using them as a direct
probe of itinerant FM in both 2D and 3D traps.  Moreover,
in the analysis of \cite{sogo2002} total magnetization is not
conserved; whereas, in most of the recent experiments
\cite{zwierlein2006, partridge2006, yshin2006}, total magnetization
is, in fact, conserved, as a consequence of isolation from the
environment and an absence of coupling between the effective spin
degree of freedom and the rest of the system. 

Ferromagnetism has been studied extensively in the context of
multi-component Bose-Einstein condensates (spinor BECs), primarily in
two dimensions, starting with \cite{ho1998, ohmi1998, battye2002,
  saito2005, leanhardt2003, sadler2006}.  However, magnetic order in
these systems is a secondary consequence of condensation, and develops
wherever the condensate forms.  Fermions near the point of degeneracy,
on the other hand, become FM in the absence of any other broken
symmetry.  Trapping potentials give rise to spatial density
variations, so different regions of the fermionic fluid at or near
degeneracy may not reach the Stoner instability\cite{stoner1938}
simultaneously, resulting in separation between the FM and
paramagnetic (PM) phases\cite{salasnich2000}, something intrinsically
absent in spinor BECs and bulk materials. This phase separation
ensures that the resulting ferromagnetic droplets have definite size,
and this, in turn, allows one to measure the phase diagram, as
discussed below.

We consider an optical lattice containing fermionic atoms cooled to
temperatures close to degeneracy. The role of spin is played by two
states in the atoms' hyperfine multiplet. Hopping between lattice
sites is determined by the overlap of the atomic wavefunctions on
those sites, while the on-site interaction term is given by the
$s$-wave scattering of ``electrons'' with opposite ``spin''
\cite{bloch2008}. In a trapped gas of equal spin populations, ground
states are constrained by vanishing magnetization: $\vec{{\cal
M}}=\int d\br\: \vec{M}(\br) = 0$.

What kind of ferromagnetic quantum state satisfies this condition?
One possibility is the state $\vert \psi \rangle =\vert S,0\rangle $
with maximal total ${\vec{\cal M}}^{2}=S (S+1)$, but ${\cal M}_{z}=0$.
The expectation values of ${\cal M}_{x}$ and ${\cal M}_{y}$ in this
state are zero, ostensibly satisfying the requirement that the total
magnetization $\langle \psi \vert \vec{\cal M}\vert \psi \rangle =0$.
However, we can rule out such a state by observing that each component
of the total magnetization operator is conserved, $[H,{\cal M}_{\alpha
}]=0$, so that the quadrupole operator $\mathcal{Q}_{\alpha \beta }=
{\cal M}_{\alpha }{\cal M}_{\beta }- \frac{1}{3}{\cal M}^{2}
\delta_{\alpha \beta }$ is also conserved: $[H,\mathcal{Q}_{\alpha
  \beta }]=0$. The state $\vert \psi \rangle $ has a non-vanishing
quadrupole moment, where $\langle \mathcal{Q}_{xx}\rangle = \langle
\mathcal{Q}_{yy}\rangle =-2\langle \mathcal{Q}\rangle_{zz}= S(S+1)/6$,
and therefore cannot develop as a spin-isotropic paramagnetic state is
cooled through the ferromagnetic transition.

\section{Landau Ginzburg Treatment}\label{}\label{}

To go further, we need to consider states of non-uniform magnetization
in which $\vec{\cal M}=0$.  To this end, we make a long-wavelength
expansion of the total energy as a functional of the local
magnetization.  Such a long-wavelength treatment of the problem does
not imply that the underlying nature of the system is classical.
Indeed, long-wavelength actions of this sort have been used to great
success in the analysis of one-dimensional quantum antiferromagnets,
where long-range order is completely absent\cite{haldane83}.

The Landau-Ginzburg free energy functional which describes
long-wavelength configurations of the magnetization  order parameter
takes the form
\begin{align}
F_{\text{LG}} 
=&\int d \br \: \frac{\rho}{2} |\nabla \vec{M}|^2 + \frac{\beta}{4}
\Bigr(|\vec{M}|^2 + \frac{\alpha(\br)}{\beta}\Bigl)^2
\label{eq:Flg}
\end{align}
Odd-power terms are ruled out by the time-reversal symmetry of the
free energy. 
Coefficients $\rho$ and $\beta$ are assumed to be
positive and constant for simplicity, and the entire effect of the
trap potentials is in the position dependence of $\alpha(\br)$.  We
define $R_c$ such that $\alpha(\br)<0$ for $r < R_c$ and $\alpha(\br)
> 0$ for $r > R_c$. We assume that the density everywhere in the
system is less than one electron per site, and conclude with a
discussion of the remaining cases.

Clearly, whatever magnetic moment is accumulated by any one region in
the trap must be completely canceled by the rest of the trap.
The configuration which connects any two regions could either be a
domain wall or some form of a twist (either with vorticity or
without). It is the competition between these three scenarios that
determines the ground state. Due to global rotation invariance, it
suffices to consider a single representative state from each class.
\fg{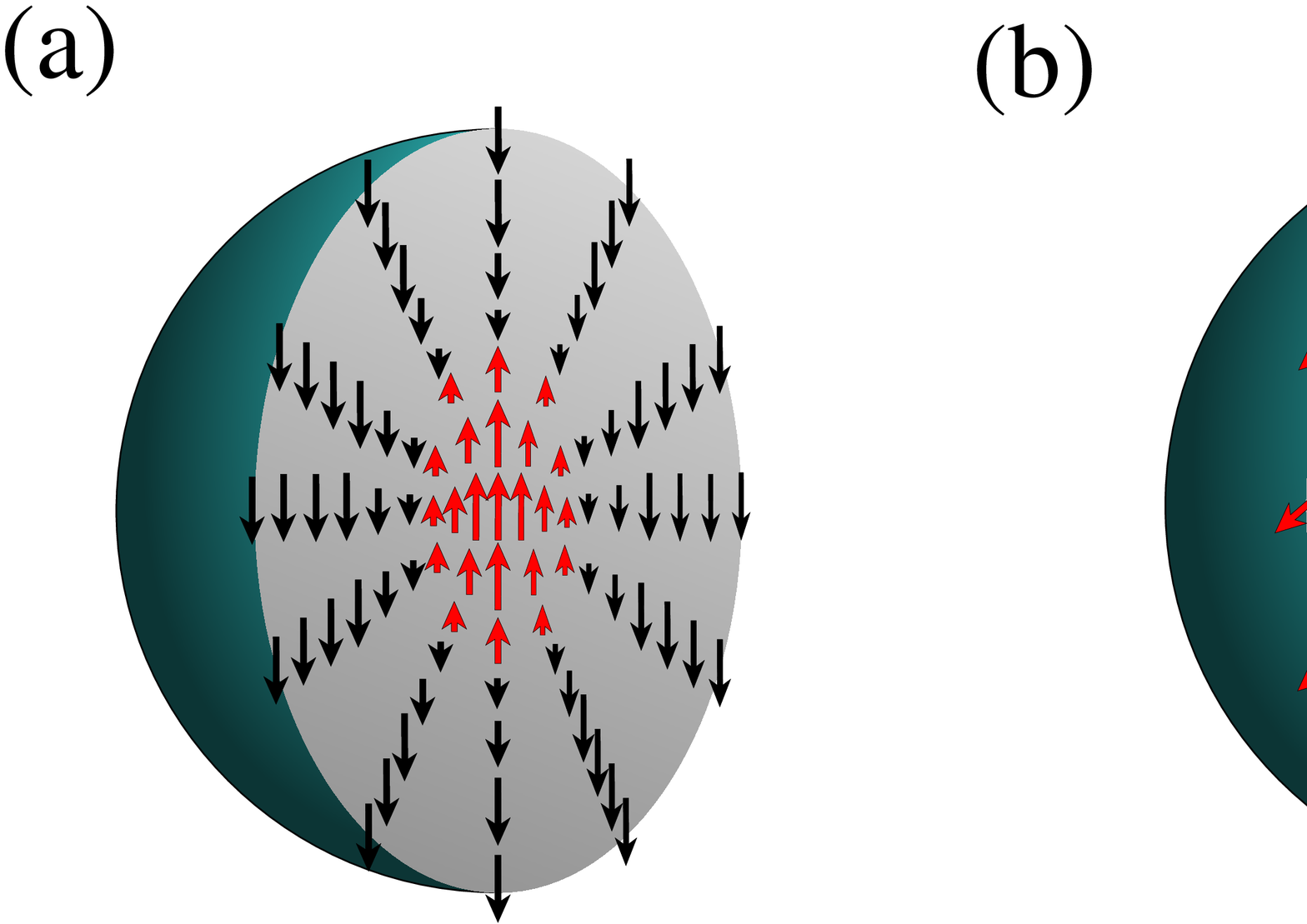}{domaintwist}{(Color online) Showing (a)
domain wall and (b) single-axis twist configuration of spins }


In 3D the first candidate is a hedgehog (Fig. \ref{hedgeskyrm} (b)).
Qualitatively, it consists of a core of radius $r_0$, where
magnetization is suppressed, a region of thickness $L$ in which it
continuously increases from $0$ to some fraction $|a| \leq 1$ of the
uniform value $M_0$, and the outer region, extending to the edge of
the trap at radius $R_T=R_c$. In a particular realization, the
magnetization vector at any point is in the radial direction. 
A competing configuration has a core
of radius $r_0$ maximally polarized in one direction. A domain wall of
thickness $L$ connects the core to the outer region polarized at a
fraction $0 \leq a \leq 1$ of the maximum $M_0$ in the opposite
direction (Fig.  \ref{domaintwist} (a)). 
The remaining possibility is a pure twist. The twist plane is globally
fixed (a rotation of $\theta_0$ about the $z$-axis), and the
magnetization turns about the $x$-axis through an angle $a\pi$, $1
\leq a \leq 2$.  The twist occupies a shell of thickness $L$ outside
of a maximally-polarized core of radius $r_0$, and the remaining outer
region is polarized along the final direction (Fig.
\ref{domaintwist}) (b)).
Parameters $a$, $r_0$, and $L$ for each configuration are determined
by numerically minimizing $F_{\text{LG}}$ for a fixed value of $R_c$
subject to the constraint of zero net magnetization.

In 2D in addition to the domain wall, the single-axis twist and the
hedgehog, which are just planar slices of their 3D cousins, there is
another possibility: the skyrmion. This configuration has a maximally
polarized core of radius $r_0$ at the center, a twist through the
angle $a\pi$, $1/2 \leq a \leq 1$ in a ring of thickness $L$, and an
outer region, polarized along the final direction (Fig.
\ref{hedgeskyrm} (a)). Unlike the single axis twist, the twist axis of
the skyrmion is a function of position.
The constraint for the 2D candidates must be
implemented explicitly, and the ground state is obtained the same way as in 3D.

Before solving the full problem, some estimates are in order. In 2D a
domain wall is approximately the suppression of the order parameter in
an annulus of thickness $L$ and inner radius $r_0$.  Omitting a
dimensionful prefactor common to all configurations, and in the case
of the domain wall, ignoring the stiffness contribution, we find the
free energy ($\xi = \sqrt{\rho/\alpha}$ is the coherence length)
\begin{equation}
  \label{eq:fdw}
  F_{\text{DW}} \sim \frac{2 r_0 L + L^2}{2\xi^2}
\end{equation}
The energy is minimized by shrinking $L$ to $L \propto \xi$, however,
the magnetization constraint forces $r_0 \propto R_c$, and so
$F_{\text{DW}} \sim R_c$. In the same units the skyrmion free energy
has contributions from vorticity and from twisting, i.e.
\begin{equation}
  \label{eq:fsk}
  F_{\text{SK}} \sim 2\ln\frac{R_c}{r_0} + (a \pi)^2 \left(\frac{r_0}{L}+\frac{1}{2}\right)  
\end{equation}
The total magnetic moment of the skyrmion is zero for some $a \sim
O(1)$. $F_{\text{SK}}$ is minimized when both $L, r_0 \propto R_c$. As
a result, the skyrmion free energy is roughly independent of the trap
size. Analysis of the single axis twist is similar, however,
implementing the constraint requires more twisting in the same volume
than in the skyrmion case, and therefore this configuration should
always be higher in energy. The 2D hedgehog free energy is estimated
to be $F_{\text{Hg}} \sim 2\ln\frac{R_c}{r_0} +
\frac{1}{2}(\frac{r_0}{\xi})^2$, sum of vorticity and core energy, and
one can show that $F_{\text{Hg}} \sim \ln{R_c/\xi}$. Thus, for large
trap sizes we expect a skyrmion to form in the trap. Similar treatment
of the 3D configurations shows that the hedgehog should be the ground
state.
\figwidth=0.8\columnwidth
\fg{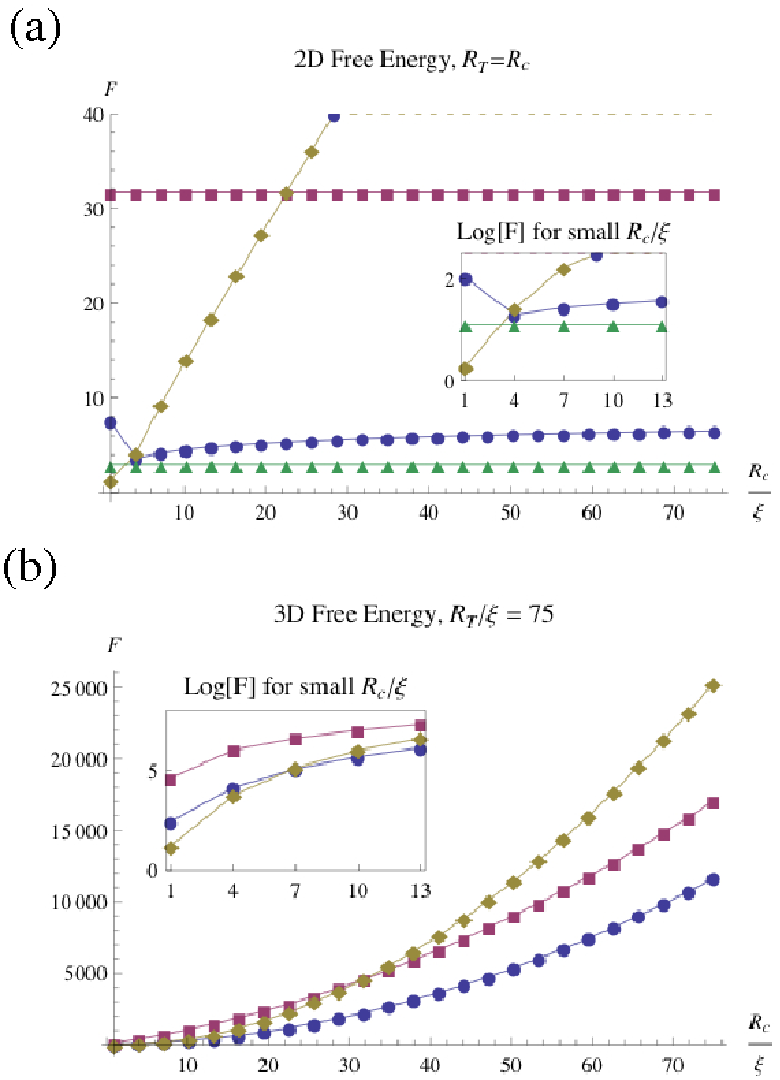}{fig:flg}{(Color online)
(a) Free energy vs. trap
    size in two dimensions, $\bullet$\ hedgehog, $\blacktriangle$ skyrmion, {\tiny
      $\blacklozenge$} domain wall, {\tiny $\blacksquare$} single axis
    twist.  Insert: $Log_{e} (F)$ vs. $R_{c}/\xi$.
(b) Free energy vs. size of the correlated region in three dimensions
for a fixed trap size, $R_T = 75 \xi$, $\bullet$ hedgehog, {\tiny
      $\blacklozenge$} domain wall, {\tiny $\blacksquare$} single axis
    twist.  Insert: $Log_{e} (F)$ vs. $R_{c}/\xi$.
}
%
Numerical computations confirm these estimates, as shown in Fig.
\ref{fig:flg}. We find that in small traps in both 2D and 3D domain
walls are preferred, while for large traps in 3D ($R_c/\xi > 7$) the
hedgehog has the lowest energy, and in 2D ($R_c/\xi > 4$), the skyrmion is
the ground state.

In the above derivation we tacitly assumed that that the whole system
becomes FM, i.e. $R_c$ = $R_T$. This is close to the truth in 2D,
since the density of states is practically flat, and the Stoner
criterion for magnetic instability dictates that the entire system
becomes FM at once, independent of the spatial density variations. In
3D, however, the density of states is more complicated, thus certain
parts of the trap will cross into the broken symmetry regime earlier
than others.  In the language of Landau-Ginzburg, it means that
$\alpha$ changes sign from positive outside of some radius $R_c$, to
negative inside.  Incorporating this into our analysis requires
somehow suppressing the magnetization outside $R_c$ in the candidate
states.  This can be accomplished by half of a domain wall, e.g.
$\vec{M}(r)$ diminishing to $0$ within some shell of thickness $L_1$
of the critical radius $R_c$ and remaining $0$ to the edge of the trap $R_T$.
Although the qualitative results of the calculations remain the same,
the numerical problem itself is modified: we minimize $F_\text{LG}$
with respect to parameters $a$, $r_0$, $L$, and $L_1$ for a given
radius of the FM region ($R_c$), and a given radius of the trap
($R_T$). Results for 3D in Fig. \ref{fig:flg} properly reflect these
considerations.

In the case when the trap has regions with density higher than one
electron per site, we expect further phase separation. In these
regions the chemical potential may enter the Mott gap and we would
expect an anti-ferromagnetic Mott phase to set in with filling locked
to one electron per site. In regions of even higher density, the
chemical potential might cross the gap and emerge, once again, in an
itinerant band. Thus, we expect the phase profile to have wedding cake
structure \cite{diener2007}. For traps with very high density in the
middle, we expect a hole PM at the core, followed in turn by shells of
hole FM, an anti-ferromagnetic Mott insulator, and an electron FM,
finally ending with the electron PM exterior. For intermediate
densities, the Mott insulator could form a natural core for the 3D
hedgehog configurations.

\section{Discussion}\label{}

\begin{figure}[t]
\vskip 0.4truein
  \begin{tabular}{ccc}
    \includegraphics[width=0.45\columnwidth]{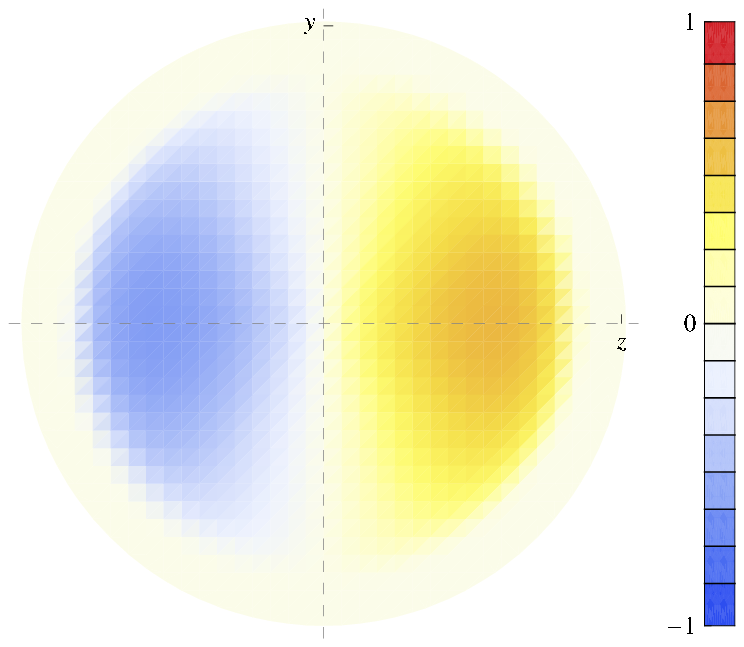}&  &
    \includegraphics[width=0.45\columnwidth]{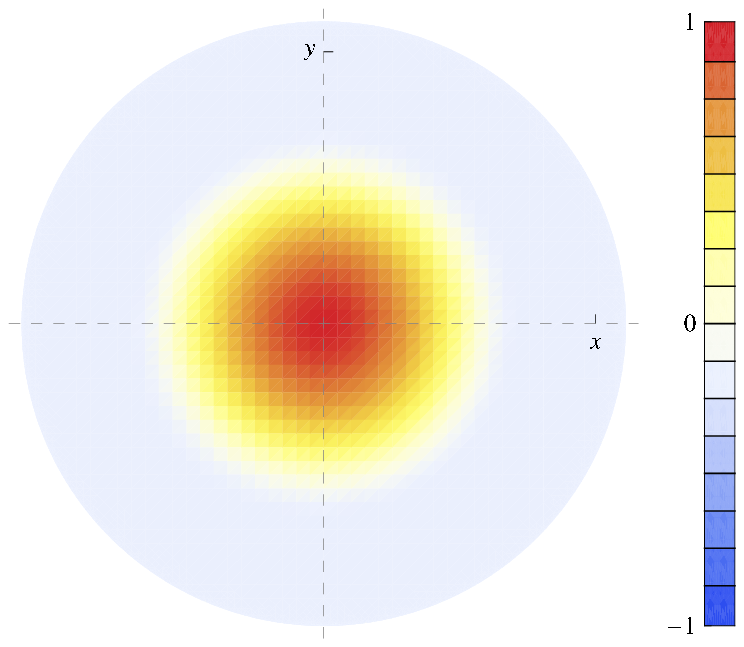}\\
  \end{tabular}
  \caption{\label{fig:model_proj}\small (Color online) False color plot of
    magnetization integrated in-situ, symmetry broken the along
    $z$-axis. Hedgehog imaged along the $x$-axis (left) and
    skyrmion imaged along the $z$-axis (right).}
  \vspace{-1.1\baselineskip}
\end{figure}

Experiments with ultra-cold atomic gases generally exploit absorption
imaging to detect the state of matter in the trap. For multi-component
gases in-situ imaging of each individual component is
possible\cite{zwierlein2006, partridge2006, yshin2006, yshin2007arx}.
In the case of FM systems this technique allows resolving the full
magnetization integrated along the camera axis. Fig.
\ref{fig:model_proj} shows in-situ images of integrated model
magnetization for the hedgehog and the skyrmion. The critical radius
can be clearly seen (and measured) in the 3D configuration, and it
designates the phase boundary between the PM outside and the FM
inside.  In addition to the phase boundary and the profile of the
magnetization, one could look for signatures of correlation in the
shot noise\cite{altman2004,trom2006}. The auto-correlator $\langle
M_z(\br) M_z(\br')\rangle$ can be calculated using the time-of-flight
images, since it is related to an integral of the frequency space
magnetization correlator. Tracking the associated correlation length,
which will diverge as the system approaches the transition, is,
potentially, a more sensitive tool and should indicate the proximity
of the FM phase before the image itself shows the phase
separation\cite{sadler2006}.

With the newly found signatures of correlation we can map out the
phase diagram of an interacting fermion gas. The density of particles
at the critical radius $R_c$ can be obtained from the absorption
images. Properly normalized, e.g. the Mott insulator is exactly at
half-filling, this density defines the critical doping of the system.
Interaction strength $U/t \sim a_s \exp(2\sqrt{V_0/E_r})$, is given by
the $s$-wave scattering length, $a_s$ and the lattice depth $V_0$,
with $a_s$ and $V_0$ (i.e. $U$ and $t$) controlled
independently\cite{bloch2008}. Thus scanning $a_s$ and $V_0$ and
measuring the critical doping for each, we determine the phase
diagram. As discussed above, at higher densities the magnetization
images should exhibit an even richer shell structure. However, by
particle-hole symmetry all of the FM regions should have the same
basic features, and each phase boundary present in the image, will
provide a point on the full phase diagram.  The finer details of the
observed configuration can give us even more details about the
parameters of the effective Landau-Ginzburg description; and in turn,
our observation of the Landau-Ginzburg parameters can be used to
evaluate the efficacy of various microscopic calculational schemes.
Similarly, we could explore the vicinity of the FM quantum critical
point, and compare the observations to the Hertz-Millis theory.
In any real experiment a number of practical issues will play a role.
In particular, the details of symmetry breaking must be the same in
successive experimental runs and in the different planes of a quasi-2D
trap configuration. One way to ensure this is to impose a small
position dependent external field. A weak interplanar coupling in
quasi-2D lattice might also stabilize the long range order in the
trap. Our model neglects the fact that in experiments to date, strong
interactions mix the various bands of the optical lattice
\cite{koehl2005,diener2006}. Elimination of this effect may require a
lower interaction strength, which in turn will require a still lower
temperature \cite{bloch2008}.

In this paper we demonstrated that ground states of interacting
fermions in both 2D and 3D traps are configurations with non-trivial
topologies.  We've determined detectable signatures of these states,
and proposed an experiment to map the phase diagram using these
signatures. Such experiments offer the prospect of directly probing
itinerant FM, studying the Stoner criterion in strongly correlated
magnets, and providing much needed experimental input in the debate on
the phase diagram of the Hubbard model.

The authors would like to thank T.-L. Ho and T. Esslinger for fruitful
discussions. Supported by NSF Grants No. DMR-0605935, DOE Grant No.
DE-FE02-00ER45790, and the Rutgers-Lucent Foundation.


\end{document}